\documentclass{appolb}
\usepackage{epsfig}

\begin{document}
\title{\vspace*{-0.14cm}Information Geometry, One, Two, Three (and Four)
\thanks{Workshop on Random Geometries, Krakow 2003.}
}
\author{D.A. Johnston
\address{School of Mathematical and Computer Sciences,\\
Heriot-Watt University,\\ Riccarton, Edinburgh EH14 4AS, Scotland}
\address{}
W. Janke
\address{Institut f\"ur Theoretische Physik,\\
 Universit\"at Leipzig,\\
Augustusplatz 10/11,
D-04109 Leipzig, Germany}
\address{}
R. Kenna
\address{School of Mathematical and Information Sciences,\\
Coventry University,\\
Coventry, CV1 5FB, England}
}
\maketitle

\begin{abstract}
Although the notion of entropy lies at the core of statistical mechanics, it is not often
used in statistical mechanical models to characterize phase transitions, a role more usually
played by quantities such as various order parameters, specific heats or susceptibilities.
The relative entropy induces a metric, the so-called information or Fisher-Rao metric, on the space 
of parameters 
and the geometrical invariants of this metric carry information about the
phase structure of the model.

In various 
models the scalar curvature, ${\cal R}$, of the
information
metric has been found to diverge at the phase
transition point and a plausible
scaling relation postulated.  For spin models the necessity of calculating
in non-zero field has limited analytic consideration to one-dimensional,
mean-field and Bethe lattice Ising models.
We report on previous papers in which we
extended the list somewhat in the current note by considering the {\it one}-dimensional Potts
model, the {\it two}-dimensional Ising model coupled to two-dimensional quantum gravity and the
{\it three}-dimensional spherical model. We note that similar ideas have been applied 
to elucidate possible critical behaviour in families of black hole solutions  in {\it four} space-time dimensions.
\end{abstract}
\setcounter{section}{-1}
\section{Introduction}

The notion of entropy,
\begin{equation}
S =  - \sum_{\alpha} p_{\alpha} \ln  \left( p_{\alpha}  \right),
\end{equation}
lies at the heart of statistical mechanics
where $p_{\alpha}$ is the probability of being in state $\alpha$.
It gives
a measure of the number of  microstates $\Omega$ accessible to a system
$S = \ln \Omega$. It is also possible to define  a relative entropy with 
respect
to some reference set of configurations of probability $r_{\alpha}$
(such as those at the critical point or zero temperature),
\begin{equation}
G ( p |  r)  = - \sum_{\alpha} p_{\alpha} \ln  \left({ p_{\alpha} \over r_{\alpha} } \right) .
\end{equation}
If $\theta$ represents parameters characterizing the class of models
under consideration, the relative entropy,
$G$, induces a metric for two ``close'' configurations
$\theta $ and $ \theta+ \delta \theta $:
\begin{equation}
dl^2 =  G \left( p ( \theta ) |  p ( \theta + \delta \theta ) \right).
\end{equation}
Since $G ( p | p ) = 0$ and $\partial G / \partial \theta = 0$
we find
\begin{equation}
dl^2 =  { \partial^2 G \over \partial \theta_i \partial \theta_j } d \theta_i
d \theta_j.
\end{equation}
If we assume a Gibbs-type distribution for the $p_{\alpha}$
\begin{equation}
p_{\alpha} = {1 \over Z } \exp ( - \theta_i \Phi_i ) \; \longrightarrow
S = \langle \theta_i \Phi_i \rangle + \ln Z,
\end{equation}
then the metric may be written as 
\begin{equation}
dl^2 = { \partial^2 \ln Z  \over \partial \theta_i \partial \theta_j } d \theta_i
d \theta_j.
\end{equation}
This metric is well known in statistics
as the Fisher-Rao metric, and characterizes the ``closeness'' of 
probability distributions \cite{Fish}. 
For a one-parameter distribution the
expectation value is the Fisher Information, $I ( \theta)$,
given by
\begin{equation}
I(\theta) = \left< - {\partial^2 \over \partial \theta^2} \ln Z \right>,
\end{equation}
and is inversely related to the variance
$I(\theta) {\rm{Var}} (\theta) =1$. In the multiparameter case this generalizes
in the obvious manner to the covariance matrix. 

Various authors
\cite{Rupe, Jany, Brody} have observed that
the geometric invariants of the metric, in particular the curvature, 
${\cal{R}}$,
might be used to characterize the phase structure of statistical mechanical models. The first observation to note is that a non-interacting model, the ideal gas, has a flat geometry since
$
{\cal{R}}_{\rm{ideal}} = 0
$. The suggestion was then that the curvature was a measure of the strength of interaction, an observation confirmed by a calculation in the case of a 
van der Waals gas (which is interacting):
\begin{equation}
R_{\rm{vdW}} = \frac{4}{3} \frac{\alpha \beta} {\bar v} {F ( \alpha , \beta
  ) \over D ( \alpha , \beta)^2},
\end{equation}
where
$\beta$ is the inverse temperature, $\alpha$ the pressure, and $\bar v(\alpha , \beta)$
and $F (\alpha , \beta)$ are expressions appearing in the
equation of state. This shows that the curvature diverges along the spinodal line where $ D ( \alpha , \beta)=0$. 

These examples suggest that phase transitions are manifested as divergences
in the scalar curvature associated with the Fisher-Rao metric, at least in two-parameter models. A natural question to ask is whether any of the standard scaling exponents associated with a transition can then be extracted from the
behaviour of the curvature. This is best pursued by consideration of some
specific examples. 
If we consider a spin model in field, the manifold of parameters, ${\cal M}$,
is two-dimensional and parametrised by
$(\theta^{1},\theta^{2})=(\beta,h)$. In this case, the components of
the  Fisher-Rao metric take the particularly 
simple form 
\begin{equation}
G_{ij} = \partial_{i}\partial_{j} f ,
\label{frmetric}
\end{equation}
where $f$ is the reduced free energy per site and
$\partial_{i} = \partial/\partial\theta^{i}$. 
With the metric given in equ.~(\ref{frmetric}),
${\cal R}$ may be calculated succinctly as
\begin{equation} 
{\cal R}\ = - \frac{1}{2 G^{2}} 
\left| \begin{array}{lll} 
\partial^{2}_{\beta} f & \partial_{\beta}\partial_{h}f & 
\partial_{h}^{2}f \\ 
\partial^{3}_{\beta}f & \partial_{\beta}^{2}\partial_{h}f & 
\partial_{\beta}\partial_{h}^{2}f \\ 
\partial^{2}_{\beta}\partial_{h}f & 
\partial_{\beta}\partial_{h}^{2}f & \partial_{h}^{3}f  
\end{array} \right| \ , 
\label{equcurv} 
\end{equation}
where $G={\rm det}(G_{ij})$. Since the only scale present near criticality for
a model displaying a higher-order transition is the correlation length,
$\xi$,
it has been hypothesised on dimensional grounds that ${\cal R} \sim \xi^d$,
where $d$ is the dimensionality of the system \cite{Rupe, Jany, Brody}.
If we assume that
hyperscaling holds, $\nu d = 2 - \alpha$,  this leads to
\begin{equation}
{\cal R} \sim |\xi|^{ ( 2 - \alpha)/\nu}.
\label{equscal}
\end{equation}

To test the behaviour of ${\cal R}$, one requires models which are
solvable in field in order to obtain analytic expressions, and these are rather
thin on the ground. Indeed, ${\cal R}$ has been calculated for the mean-field
and
Bethe-lattice
Ising models \cite{Brian} and the above scaling behaviour verified. It has also
been calculated for the one-dimensional Ising model \cite{Jany} where it takes
the remarkably simple form,
\begin{equation}
{\cal R}_{\rm{Ising}}=1 + {\cosh h \over \sqrt{\sinh^{2}h+e^{-4\beta}}}.
\label{RIsing}
\end{equation}
In this case ${\cal R}$ is positive definite
and diverges only at the zero-temperature, 
zero-field
``critical point'' 
of the model. The correlation length
is given by
\begin{equation}
 \xi^{-1} = -\ln{\left(\tanh (\beta)\right)},
\end{equation}
so that $\xi \sim  \exp(2\beta)$ near criticality, 
and (\ref{equscal}) holds there with $\alpha=1, \nu=1$ as 
expected\footnote{The Bethe lattice model
also satisfies the postulated scaling, although there are some subtleties
coming from the exponent $\alpha$ being zero \cite{Brian}.}.

Given this rather short list of examples, any further additions 
would be very worthwhile in order to see which features 
are generic and which are particular to the models concerned.  
We expand the list incrementally here, looking at the 
{\it one}-dimensional Potts model, the {\it two}-dimensional Ising model
coupled to two-dimensional quantum gravity and 
the {\it three}-dimensional spherical model, which shares the critical exponents of the just mentioned coupled Ising model. 
We also discuss the application of similar ideas to critical behaviour in various families of black holes. The present paper is essentially an amalgam
of results presented in \cite{no1,no2,no3}, with some added discussion
of related work in black hole physics.


\section{One}

The partition function for the 1D $q$-state Potts 
model is given by
\begin{equation}
Z_N (y, z) =\sum_{\{\sigma\}} \exp \left[ {\beta \sum_{j=1}^N \left( \delta( \sigma_j,  \sigma_{j+1})  - {1 \over q} \right) +  h
\sum_{j=1}^N \left( \delta ( \sigma_j, 1) - {1 \over q} \right)} \right]
,
\label{ZPotts}
\end{equation}
where the spins, $\sigma_j\in \{1,\dots, q\}$, are defined on the
sites, $j \in \{1,\dots N\}$, of the lattice
and where
we have defined $y = \exp ( \beta )$ and $z = \exp ( h )$ for later calculational convenience.
The model may be solved by transfer matrix methods \cite{Glue}, just as the 1D Ising model. 
For general $q$ the full transfer matrix $T ( y, z)$ 
may be written as $q-2$ diagonal elements, 
$( y - 1) (  y z )^{-1/q}$, and
a $2 \times 2$ factor $t ( y, z)$:
\begin{equation}
t ( y, z) = {1 \over ( y z )^{1/q}  }\pmatrix{
 y z & z^{1/2} \sqrt{ q - 1} \cr
 z^{1/2}\sqrt{q - 1}  & y + q - 2 \cr}.
\label{eP}
\end{equation}
The partition function is $Z_N (y, z) = {\rm Tr}\, T (y, z)^N$ 
and the eigenvalues of 
$T(y,z)$ are $\lambda_0,\lambda_1,\dots,\lambda_{q-1}$, where
\begin{equation}
\left.
\begin{array}{ll}
\lambda_{0}\\
\lambda_{1}
\end{array}
\right\}
 = {1 \over 2} \left[{  y ( 1 +z ) + q - 2    
\pm 
\sqrt{ (y ( 1 - z) + q - 2)^2 + ( q- 1) 4 z } \, }\right] \, ( y z ) ^{ - {1 \over q}}
\label{eigen}
\end{equation}
and $\lambda_2, \dots \lambda_{q-1} = ( y - 1 ) (y z ) ^{ - {1 / q}}$.
The reduced free energy per site in the thermodynamic limit,
$N\rightarrow \infty$, is thus given by $f = -\ln \lambda_{0}$.

It is straightforward to use this expression for the  free energy in 
equ.~(\ref{equcurv})
to obtain the curvature, ${\cal R}$. In the current notation we re-derive the expression for the Ising model \footnote{There is a factor
of two difference in the definitions of $\beta, h$  between
the Ising and Potts notations coming from the
different spin definitions.}
as
\begin{equation}
{\cal R}_{\rm{Ising}}=1 + {y ( 1 + z) \over \sqrt{y^2 - 2 y^2 z + y^2 z^2 +4 z}}.
\end{equation}
The expression for general $q$ is similar in form to this, and is
\begin{equation}
{\cal R}_{\rm{Potts}} = A (q, y, z)   + {B(q, y , z) \over  \sqrt{\eta (q, y, z)}},
\end{equation}
where the coefficients 
$A (q, y, z)$ and $B(q, y, z)$
are smooth functions of $y$ and $z$ and do not diverge for finite (physical) temperature
or field. Furthermore
\begin{equation}
\eta (q, y, z) =  [ y ( 1 - z) + q - 2 ]^2 + ( q- 1) 4 z .
\label{eta}
\end{equation}
The expressions for $A(q, y, z)$ and $B(q, y , z)$ are very lengthy for general $q$ (although still easily obtained)
so we do not reproduce them here.

In zero-field ($z=1$) the 
expression for ${\cal R}$  is much more compact
and is  written for general $q$ as
\begin{equation}
{\cal R}  = { (y + q -1 ) ( 4 y^2 + ( q - 2 ) y - (q - 2) ( q  - 1)) \over (q - 1) ( 2 y + q  - 2 )^2 }.
\label{Rz1}
\end{equation}
We see that as $y$ ranges from $1$ to $\infty$, ${\cal R}$ ranges
from $(4 - q) / ( q -1)$ to $\infty$. In particular, the sign of the
$y=1$ ($\beta=0$) limit of ${\cal R}$ changes at $q=4$, although
the general morphology of ${\cal R}$ 
as  a function of $y$ and $z$ remains the same for all $q>2$
as we shall see below.

The correlation length for the one-dimensional Potts model
is defined in a similar manner to that of the Ising model,
\begin{equation}
 \xi^{-1} = -\ln{\left(  {\lambda_1 \over \lambda_0} \right)},
\end{equation}
so $\xi \sim y$ for $z=1, y \rightarrow \infty$. 
We thus retrieve the scaling  of
${\cal R}$ 
for the one-dimensional
Potts model
expected from equ.~(\ref{equscal}), 
namely ${\cal R} \sim y$ as $y \rightarrow \infty$. 
The exponents, as for the Ising model,
are $\alpha=1, \nu=1$.

The general features of ${\cal R}$ at non-zero temperature and field
are perhaps easiest seen in a 
contour plot as a function of $y$ and $z$. In Fig.~1 we show the Ising ($q=2$) case
which has certain non-generic features. The  $\pm h$ symmetry
of the Ising model appears as a $z \rightarrow 1/z$ symmetry
in the plot of ${\cal R}$. In addition, one can see that ${\cal R}$
is positive for all $y$ and $z$. The maximum of ${\cal R}$ 
for a given $y$ value lies
along the zero field line at $z=1$.
In Fig.~2, ${\cal R}$ is plotted for the $3$-state 
Potts model, 
We see that there is no longer a $z \rightarrow 1/z$ symmetry
and that ${\cal R}$  is no longer   positive definite. 
 
Some years ago Lee and Yang \cite{YL} 
addressed the question of how the singularities
associated with field-driven phase transitions in Ising-like spin models 
on lattices
arose in the thermodynamic limit. 
This was later extended by various authors 
to other models and to temperature-driven transitions \cite{others,fishZ}.
Lee and Yang
observed that the zeroes of the partition function
for a spin model in a complex external field
on a finite lattice would give rise
to singularities in the free energy. 
In the thermodynamic limit 
these complex zeroes
move in to pinch the real axis, signalling the onset of a
physical
phase transition.
Typically, the loci of zeroes are lines in the complex field
or temperature plane
and when the
endpoints of such lines
occur at non-physical (i.e.\ complex) external field values they can
be considered as ordinary critical points with an associated edge critical
exponent, usually dubbed the Lee-Yang edge exponent \cite{others}. 

The Lee-Yang zeroes for the one-dimensional
Potts model on a periodic chain with $N$ sites
are given by the solutions \cite{Glue, Brian2} of
\begin{equation}
Z_N = (\lambda_1)^N +(\lambda_0)^N = 0 \qquad\Leftrightarrow \qquad
\lambda_1 = \exp{\left( {in\pi\over N} \right)}
\lambda_0    
\end{equation}
where $\lambda_{0,1}$  are the eigenvalues given in 
equ.~(\ref{eigen}) and $n$ is odd. In the thermodynamic limit
the locus of zeroes is determined by $| \lambda_0 | = | \lambda_1 |$
or
\begin{equation}
\eta(q, y, z) = [y ( 1 - z) + q - 2]^2 + ( q- 1) 4 z  = 0,
\end{equation}
which can be satisfied for complex (in
the $q=2$ Ising case, purely imaginary) values of  $h$.

From these considerations, 
it is clear that
${\cal R}$ will also diverge as
the locus of zeroes is approached 
for both Ising and Potts
models if we allow ourselves the liberty of an analytic
continuation of the field to complex $h$ values once 
${\cal R}$
has been calculated,
since ${\cal R} = A + B / \sqrt{\eta}$ 
and $A, B$ are  finite as $\eta \rightarrow 0$.
The presence of the square root means that the  
divergence is characterised by an exponent $\sigma=-1/2$ which is 
the Lee-Yang edge exponent for the one-dimensional
Potts (and Ising) model \cite{others}. 

The status of these
observations is a little unclear to us, since the calculation
of ${\cal R}$ has assumed a real metric geometry throughout and such
an arbitrary continuation in the final
expression might be rather dangerous. 
It is nonetheless interesting that the Lee-Yang edge transition 
is still visible as a divergence in ${\cal R}$. 

\section{Two}

The solution of the Ising model on 
an ensemble of $\Phi^4$ ($4$-regular) or $\Phi^3$
($3$-regular)
planar random graphs was first presented  
by Boulatov and Kazakov \cite{BK},
who were motivated by string-theoretic
considerations, since the 
continuum limit of the theory represents
matter coupled to 2D quantum gravity. They considered the partition function
for the Ising model on a single $n$ vertex planar graph 
with connectivity matrix $ G_{ij}^n $:
\begin{equation}
Z_{{\rm single}}(G^n,\beta,h) =
\sum_{\{\sigma\}} \exp \left({\beta}\sum_{\langle i,j \rangle} G^n_{ij}\sigma_i
\sigma_j + h \sum_i \sigma_i\right)\, ,
\end{equation} 
then summed it over all $n$ vertex graphs $\{G^n\}$ 
resulting in
\begin{equation}
Z_n = \sum_{\{G^n\}} Z_{{\rm single}}(G^n,\beta,h)\ ,
\end{equation} 
before finally forming the grand-canonical sum over 
graphs with different numbers $n$ of
vertices 
\begin{equation}
{\cal Z} = \sum_{n=1}^{\infty} \left( - 4 g c \over 
( 1 - c^2 )^2 \right)^n Z_n\ ,
\label{grand}
\end{equation}
where $c = \exp ( - 2 \beta)$.   
This last expression
could be calculated exactly as matrix integral
over $N \times N$ Hermitian matrices,
\begin{eqnarray}
\lefteqn{
{\cal Z} 
 = - 
 \ln 
 \int {\cal D}
 \phi_1~{\cal D}\phi_2~ 
\exp 
\left( 
       -{\rm Tr}\left[
                     {1\over 2}(\phi_1^2+\phi_2^2)
               \right.
\right.
}
& &
\nonumber
\\
& &
\quad \quad \quad \quad \quad \quad \quad \quad \quad \quad \quad \quad
\left. \left.
- c \phi_1\phi_2  - 
\frac{g}{4}( \e^h \phi_1^4 + \e^{-h} \phi_2^4)\right]  \right),
\label{matint}
\end{eqnarray}
where the $N \to \infty$ limit is to be taken to pick out the planar diagrams
and the potential appropriate for $\Phi^4$ (4-regular) random 
graphs has been shown.

When the matrix integral is carried out the solution is given 
in parametric form by
\begin{equation}
{\cal Z} = {1\over 2}\ln \left({z \over g}\right)-{1\over g}\int_0^z~{dt\over t}g(t)
+{1\over 2g^2}\int_0^z{dt\over t}g(t)^2,
\label{fullpart}
\end{equation}
where the function $g(z)$ is 
\begin{equation}
g(z)=\frac{1}{9} c^2 z^3 +  \frac{z}{3}
\left[ \frac{1}{(1- z)^2} - c^2 +\frac{z B}{(1- z^2)^2}
\right]
\label{geq}
\end{equation}
and $B= 2 [ \cosh ( h ) - 1]$.

In the thermodynamic limit the free energy per site is given by 
\begin{equation}
f  
=  \ln \left( {-4 c g (z_0) \over ( 1 - c^2)^2} \right)\ ,
\label{equfL}
\end{equation}
where $z_0 = z_0 ( \beta, h)$ is the appropriate low- or high-temperature 
solution of
$ g' ( z ) =0$. When $h=0$ this may be solved in closed form, and although the
solution is not available explicitly
for non-zero $h$ it can still be developed
as a power series in $h$ around the zero-field solutions in order to 
obtain expressions for quantities such as the energy, specific \ heat,
magnetization and
susceptibility. It was found that
the critical exponents were given by $\alpha=-1$, $\beta=1/2$, $\gamma=2$,
so the transition was {\it third} order with, intriguingly, the same
exponents as the 3D spherical model on a regular lattice \cite{Stau}.

If we carry out a perturbative expansion
for the high-temperature solution,
which is symmetric in $h$ and hence a series in even powers, we
find
\begin{eqnarray}
\lefteqn{
z_{0} =
1-\frac{1}{u}-\frac{(u-1)(2 u^2-2 u+1)}{(2 u-1)^4} h^2 
}
& & 
\nonumber \\
& + &  \frac{(u-1) (2 u^2-2 u+1) ( 4 u^5-10 u^4+10 u^3-5 u^2+5 u+1)}{24 (2 u-1)^9} h^4 + \dots\ ,
\quad \quad
\label{equz0}
\end{eqnarray}
where the coefficients
in the series are
 most naturally expressed in terms of $u = \exp ( - \beta)
= \sqrt{c}$, as above.

The expected scaling form of the various components
of ${\cal R}$ for a 
generic spin model in field is discussed at some length in \cite{jany},
and we now recapitulate these results briefly for comparison with
the specific results for the Ising model on planar random graphs.
The starting point is the scaling form
of the free energy per spin near the critical point,
\begin{equation}
f(\epsilon, h) = \lambda^{-1} f ( \epsilon \lambda^{a_{\epsilon}} , 
h \lambda^{a_h} )\ ,
\end{equation}
where $\epsilon \equiv \beta_c - \beta$ and $a_{\epsilon}, a_h$
are the scaling dimensions for the energy and spin operators. For
$\epsilon>0$, i.e., in the unbroken high-temperature phase, we can use 
standard scaling assumptions to write this as
\begin{equation}
f(\epsilon, h) =  \epsilon^{1 / a_{\epsilon}} \psi_{+} ( h \epsilon^{- a_{h} / a_{\epsilon}} ),
\end{equation}
where $\psi_{+}$ is a scaling function
and we also define $A = 1 / a_{\epsilon}$ and $C = - a_{h} / a_{\epsilon}$
for later convenience.  In terms of the standard critical exponents
$A = 2 - \alpha$ and $A + C = \beta$.

This generic scaling form can now be substituted into
equ.~(\ref{equcurv}) to find the scaling behaviour 
of the 
scalar curvature (\ref{equcurv}) near criticality (i.e. $h=0$, 
$\epsilon \rightarrow 0$),
\begin{equation} 
{\cal R } =  { ( A + 2 C ) [ (A + 2 C ) - ( A - 2 )] \over 2 A ( A -1 ) \psi_{+} ( 0 ) }  \epsilon^{-A}
\end{equation}
or, translating back to the standard critical exponents,
\begin{equation}
{\cal R } = { \gamma ( \gamma - \alpha ) \over 2 ( 2 -\alpha ) ( 1 - \alpha ) \psi_{+} ( 0 )}
\epsilon^{\alpha -2}\ .
\end{equation}
The discussion in \cite{jany} was intended to be as general as possible,
one should note that for Ising-like models with a $\pm h$ symmetry
all odd derivatives of the scaling function w.r.t.\ $h$ will vanish
so $ \partial^3_h f =0$ rather than $\epsilon^{A + 3 C} \psi_{+}^{'''} ( 0 )$.
This does not affect the stated scaling relations. 

However, one feature
of these scaling relations does have an impact on the observed
scaling for the Ising model. Generically one expects that
$\partial_{\beta}^2 f = A (A - 1) \epsilon^{A-2} \psi_{+} (0)$,
which contributes to both the metric and the determinant
involved in calculating ${\cal R}$. If $A>2$, i.e.\ $\alpha<0$, this
naively suggests a vanishing $\partial_{\beta}^2 f$ at
criticality, which will in general {\it not} be the case. 
There would instead be a  contribution from a regular term, which would 
give a constant at the critical point. Having such a 
constant term, which we take to be $A ( A - 1) \phi ( 0)$ for notational convenience, 
modifies the scaling form of ${\cal R}$ in the case $\alpha < 0$,
$A > 2$ to
\begin{equation}
{\cal R}\ =   { ( A + 2 C )^2 \over 2 A ( A -1 ) \phi ( 0 ) }  \epsilon^{-2}
\label{equR3}
\end{equation}
or
\begin{equation}
{\cal R } = { \gamma^2 \over 2 ( 2 -\alpha ) ( 1 - \alpha ) \phi ( 0 )}
\epsilon^{ -2}\ ,
\label{equR3a}
\end{equation}
so the critical exponent $\alpha$ no longer appears in the scaling
exponent.

By virtue of the 
Boulatov and Kazakov solution of the Ising model on planar random graphs
\cite{BK}
we can  explicitly confirm these observations.
Since $\alpha=-1, \beta = 1/2, \gamma = 2$, we have $A=3$, $C=-5/2$
and the modified discussion of scaling should apply.
We can now take the series expansion for $z_0$ from  equ.~(\ref{equz0}),
insert this into $g(z)$ and 
use the resulting expression for $f$
in  equ.~(\ref{equfL}) 
to calculate the various terms
that appear in the scalar curvature ${\cal R}$ 
as power series
in $h^2$.
We find that the leading terms at $h=0$, with $\epsilon_u \equiv u - u_{cr} =
\epsilon/2 + \ldots$
and $u_{cr} = 1/2$, and using $\beta, h$ as co-ordinates are
\begin{equation}
{\cal R} \sim \frac{225}{704 } \epsilon_u^{-2} + \ldots = \frac{225}{176}
\epsilon^{-2} + \ldots\ .
\label{finalscaling}
\end{equation}
A glance back at equ.~(\ref{equR3a}) 
shows that the modified scaling
for $A>2$ that these incorporate is, indeed, followed.

\section{Three}

Berlin and Kac \cite{BKac} introduced the spherical model
(and the Gaussian model) in an attempt to understand 
how generic some
of the features of Onsager's solution \cite{Ons} of the $2$D Ising model  
are for
ferromagnetic spin models, particularly for other dimensions.
In the spherical model, the $\pm 1$ condition on the value
of 
the Ising spins is relaxed, 
whilst retaining a global constraint on the 
{\it total} spin magnitude. With $s_i$ denoting the value of a spin
at a site $i$ of a hypercubic lattice, the partition function is \cite{BK} 
\begin{equation}
{\cal Z} = \int ds_1 \ldots ds_N \exp \left(  \beta \sum_{\langle ij \rangle}
s_i s_j  + h \sum_i s_i \right) \delta 
\left( \sum_i s_i^2 - N \right),
\label{sph}
\end{equation}
where $N$ is the total number of sites.
This can be evaluated by exponentiating the constraint and using 
steepest descent, resulting in the following expression for the 
reduced free energy
per site in the thermodynamic limit, $N \rightarrow \infty$:
\begin{equation}
f =  \frac{1}{2} 
\ln \left( { \pi \over \beta } \right) + \beta  z  
-  \frac{1}{2} g ( z ) + {h^2 \over 4 \beta (z - d )},
\label{sphsol}
\end{equation}
where 
\begin{equation}
g( z ) = \frac{1}{(2 \pi)^d}  \int_0^{2 \pi} 
d \omega_1 \ldots d \omega_d \ln 
\left( z  - \sum_{k=1}^d \cos ( \omega_k ) \right).
\label{saddlefunc}
\end{equation}
The saddle-point value of $z$ which appears in the expression for the 
free energy in equ.~(\ref{sphsol}) is determined from
\begin{equation}
g' (z) =  2 \beta - { h^2   \over 2 \beta (z - d)^2} .
\label{saddle}
\end{equation}
The solution reveals no transition for $d=1$ and $2$, 
and a transition 
with the exponents $\alpha=-1, \; \beta=1/2,
\gamma=2$ for $d=3$.

It is useful to note that, with $h=0$, equ.~(\ref{saddle}) 
gives
\begin{equation}  
{d z \over d \beta } = { 2 \over g'' ( z)} ,
\label{zscal1}
\end{equation}
and hence
\begin{equation}
{d^2 z \over d \beta^2 } = - 
{4 g^{(3)} ( z) \over [g'' ( z)]^3 } .
\label{zscal2}
\end{equation}
The critical point is given by $z = d = 3$ and $h=0$ \cite{BK}, 
and the behaviour of $g(z)$ in this 
region is  determined by
differentiating equ.~(\ref{saddlefunc}) twice and then expanding
for the small $\omega_k$ values which give the dominant contribution.
One finds
\begin{eqnarray} 
g'' ( z) \sim  -{1 \over 2 \sqrt{2} \pi} (z - 3)^{-1/2} .
\label{gscal2}
\end{eqnarray}
A further differentiation gives
\begin{eqnarray}
g^{(3)} ( z ) \sim  { 1 \over 4 \sqrt{2} \pi} (z - 3)^{-3/2}
,
\label{gscal3}
\end{eqnarray}
and an integration yields
\begin{equation}
g' (z) =  { 1 \over \sqrt{2} \pi }  (z - 3)^{1/2}  + g' (3) , 
\end{equation}
where $g'(3) = 
(18 + 12 \sqrt{2} - 10 \sqrt{3} - 7 \sqrt{6}) 
           [K(2 \sqrt{3} + \sqrt{6} - 2 \sqrt{2} - 3)]^2
\approx 0.505\,462\,019\dots$
is the massless 3D lattice propagator at the
origin, which can be expressed in terms of one of the three classical Watson
integrals \cite{3D_latt_prop} and hence is given by the standard elliptic 
integral $K(k^2)$.
This latter expression 
can be combined with equ.~(\ref{saddle}) with $h=0$
to give
\begin{equation}
(z - 3 ) \sim  8 \pi^2 ( \beta_c - \beta)^2 \sim  \epsilon^2,
\label{eqscal1}
\end{equation}
in which  $\beta_c= g' (3)/2 \approx 0.252\,731\,009\dots$.
Equations (\ref{gscal2}) and (\ref{gscal3}) may then be substituted in
equs.~(\ref{zscal1}) and (\ref{zscal2}) to 
give the scaling of $d z / d \beta$ and $d^2 z / d \beta^2$,
\begin{eqnarray}
\lim_{z \to 3} {d z \over d \beta } & = & 
\lim_{z \to 3} \{ - 4 \sqrt{2} \pi    (z - 3 )^{1/2} \} = 0 , \nonumber \\
\lim_{z \to 3} {d^2 z \over d \beta^2 } & = & 16 \pi^2 ,
\label{eqscal2}
\end{eqnarray}
which we shall employ below in the calculation of the scalar curvature.

We now move on to examine the scaling of the various 
terms contributing to ${\cal R}$ in equ.~(\ref{equcurv})
for the spherical model.
As we have remarked, the $h \to -h$ symmetry in the
free energy per site, $f$, of the spherical model means that
any terms with an odd number of $h$ derivatives will automatically
be zero when $h=0$, hence $f_{\beta h} = f_{\beta \beta h} = f_{h h h}=0$. This leaves
the non-zero terms in the scaling region
\begin{eqnarray}
f_{\beta \beta} &\sim& {1 \over 2 \beta_c^2}  , \nonumber \\
f_{h h } &\sim&  {1 \over 16 \pi^2  \beta_c (\beta_c - \beta)^{2}} 
\sim \epsilon^{-2} ,  \nonumber \\
f_{\beta \beta \beta} &\sim&  16 \pi^2 - { 1 \over  \beta_c^3}  , \nonumber \\
f_{h h \beta} &\sim&  
{1 \over 8 \pi^2 \beta_c (\beta_c - \beta)^3 } 
- {1 \over 16 \pi^2  \beta_c^2 ( \beta_c - \beta)^2   }
\sim  \epsilon^{-3} .
\end{eqnarray}
We see that the expected general scaling of each term (for
$\alpha<0$) does indeed apply and that overall we have,
as in equ.~(\ref{equR3a}),   
\begin{equation}
{\cal R} \sim  \epsilon^{-2}.
\end{equation}
We thus see that calculating the scaling of ${\cal R}$
for the $3$D spherical model for which $\alpha=-1$ gives
results in accordance with expectations from general scaling arguments
which take into account the negative $\alpha$, similarly to the Ising
model on planar random graphs.

\section{Four}

The thermodynamics of black holes has been a subject of abiding interest
since the pioneering work of Hawking \cite{Hawk} and
similar ideas to those discussed in the previous sections 
for statistical mechanical models
have also been applied to investigations of the critical behaviour of
various families of black hole solutions in general relativity.
Critical behaviour has arisen in several contexts in the study
of black holes, ranging from the Hawking-Page \cite{HP} phase transition in hot
Anti-de-Sitter space and the pioneering work by Davies \cite{Davies} on the
thermodynamics of Kerr-Newman black holes, to
the idea that the extremal limit of various black hole
families might be regarded as a bona-fide critical point \cite{Louko, Chamblin, Cai, Aman}.

It is the latter that is perhaps the closest to the work described here.
In these studies the metrics used are the
Ruppeiner metric
\begin{equation}
g_{ij} = - \partial_i \partial_j S ( E , N^a ),
\end{equation}
and the Weinhold \cite{Weinhold} metric
\begin{equation}
g_{ij} = \partial_i \partial_j E ( S , N^a ),
\end{equation}
where $S$ is the entropy and $E$ the energy, with the  $N^a$ 
being other extensive variables such as scalar charges. Both metrics
are related to the Fisher-Rao metric we have used by Legendre transforms
of the appropriate variables, and have also been employed in a statistical mechanical context.

For example \cite{Aman}, a  Reissner-Nordstr\"om black hole has a  Weinhold  metric of the form
\begin{equation}
dl_{W}^2 = {1 \over 8 S^{(3/2)}} \left( \left( 1 - {3 Q^2 \over S} \right)
  dS^2 - 8 Q dQ dS + 8 S dQ^2 \right),
\end{equation}
where $Q$ is the charge, and we can see that one of the metric components vanishes at $S = 3 Q^2$.
This led Davies \cite{Davies} to suggest that there was a phase transition at this point,
but transforming to the Ruppeiner metric and choosing new co-ordinates gives a {\it flat}
metric with $R=0$ everywhere. 

On the other hand Kerr black holes which possess a spin $J$ display a curved Ruppeiner geometry
with a scalar curvature
\begin{equation}
R = \frac{1}{ 4 M^2} { \sqrt{ 1 - {J^2 \over M^4} } -2 \over \sqrt{
    1 - {J^2 \over M^4}}},
\end{equation}
that diverges at the extremal limit
\begin{equation}
{J \over M^2} = \pm 1.
\end{equation}
It would thus appear that the general framework
of looking for the signal of phase transitions
in the geometric invariants of the information
metric (or related metrics) can also be employed
in these circumstances. Indeed, the AdS/CFT 
correspondence identifies Hawking-Page type transitions with
deconfinement in the dual gauge theories \cite{Chamblin, Clifford}
so it might be profitable to explore the information geometry
view of such transitions further.

\section{Acknowledgements}

D.J. and W.J. were partially supported by
EC IHP network
``Discrete Random Geometries: From Solid State Physics to Quantum Gravity''
{\it HPRN-CT-1999-000161}. 



\clearpage \newpage
\begin{figure}[t]
\vskip 15.0truecm
\includegraphics{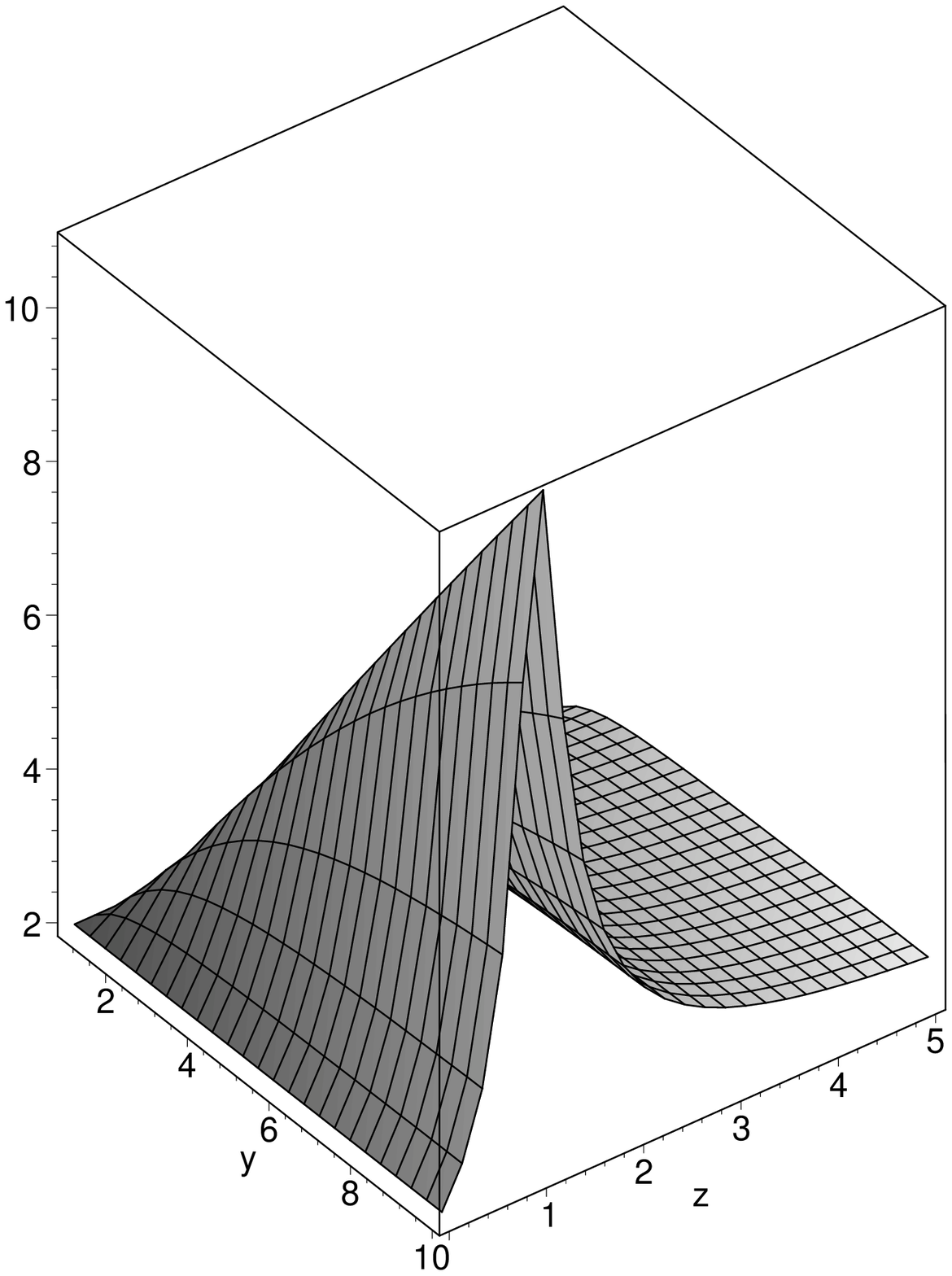}
\caption[]{\label{fig1} A plot of ${\cal R}$ for the one-dimensional Ising model
for $y=1 \ldots 10$, $z=0 \ldots 5$. The positivity of ${\cal R}$ and
the expected $z \rightarrow 1/z$ symmetry are both apparent.
}
\end{figure}


\clearpage \newpage
\begin{figure}[t]
\vskip 15.0truecm
\includegraphics{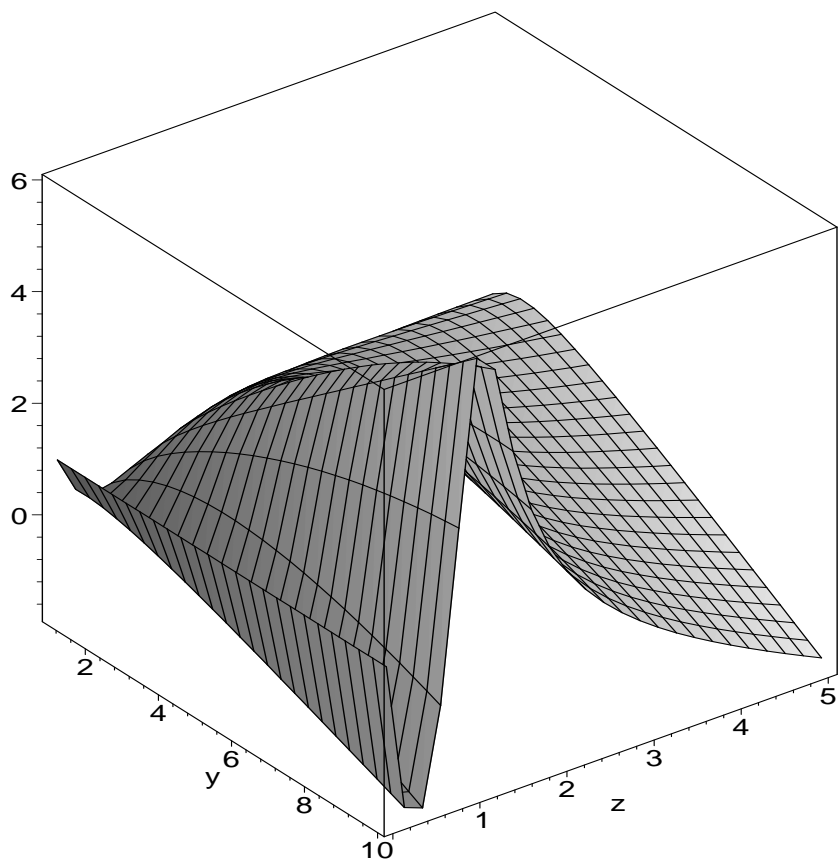}
\caption[]{\label{fig2} A plot of ${\cal R}$ for the one-dimensional 
$3$-state Potts model
for $y=1 \ldots 10$, $z=0 \ldots 5$. ${\cal R}$ is no longer
positive definite for physical values of $y,z$ and
there is no $z \rightarrow 1/z$ symmetry. In addition the maximum of
 ${\cal R}$ does not lie at $z=1$.
}
\end{figure}


\end{document}